\documentstyle[12pt]{article}
\newcommand{\drm}{{\rm d}}
\newcommand{\bi}{\bibitem}
\newcommand{\til}{\tilde}

\newcommand{\Lc}{{\cal L}}

\newcommand{\be}{\begin{equation}}
\newcommand{\ee}{\end{equation}}
\newcommand{\bea}{\begin{eqnarray}}
\newcommand{\eea}{\end{eqnarray}}
\newcommand{\beas}{\begin{eqnarray*}}
\newcommand{\eeas}{\end{eqnarray*}}

\newcommand{\ga}{\gamma}

\newcommand{\erm}{{\rm e}}
\newcommand{\gabf}{\mbox{\boldmath $\gamma$}}

\newcommand{\dar}{\dagger}

\newcommand{\sig}{\sigma}

\newcommand{\longra}{\longrightarrow}
\newcommand{\upa}{\uparrow}

\newcommand{\veps}{\varepsilon}

\newcommand{\C}{{\rm I}\!\!\!{\rm C}}
\newcommand{\La}{\Lambda}

\newcommand{\lan}{\langle}

\newcommand{\ran}{\rangle}

\newcommand{\R}{{\rm I}\!{\rm R}}

\newcommand{\h}{\hspace*{0.5 cm}}
\newcommand{\dis}{\displaystyle}

\newcommand{\psit}{\tilde{\psi}}
\newcommand{\bt}{\beta}

\newcommand{\cent}{\centerline}
\newcommand{\vs}{\vspace*}

\newcommand{\r}{\rho}
\newcommand{\dpar}{\partial}
\newcommand{\pa}{\dpar}

\newcommand{\dox}{\dot{x}}
\newcommand{\dopi}{\dot{\pi}}
\newcommand{\dopsi}{\dot{\psi}}

\begin{document}
 
\cent{{\bf ABOUT ZITTERBEWEGUNG AND ELECTRON STRUCTURE$^{\: \dag}$}}

\vs{1 cm} 

\cent{{Waldyr A. RODRIGUES Jr., \ \ Jayme VAZ Jr.}}

\vs{0.3 cm}
 
\h\h\h {\em Departamento de Matem\'{a}tica Aplicada,\\
\h\h\h\h Universidade Estadual de Campinas; 13083-970 Campinas, S.P.;
Brazil}.\\
 
\vs{0.1 cm}

\cent{ and }

\vs{0.3cm}

\cent{{Erasmo RECAMI$^{\, (*)}$, \ \ Giovanni SALESI}}

\vs{0.3 cm}

\h\h\h {\em INFN -- Sezione di Milano, 57 Corsitalia, Milan,
Italy.} 

\footnotetext{$^{\: \dag}$ Work partially supported by INFN, CNR, MURST
and by CAPES, CNPq, FAPESP.} 
\footnotetext{$^{\, (*)}$ Also: Facolt\`{a} di Ingegneria, Universit\`{a} 
Statale di Bergamo, Dalmine (BG), Italy; \ and  C.C.S.;
State University at Campinas; 13083--Campinas, S.P.; Brazil.} 

\vs{2cm}
 
{\small\bf ABSTRACT:}  We start from the spinning electron theory by
Barut and Zanghi, which has been recently translated into the Clifford algebra
language.  We ``complete" such a translation, first of all, by expressing in 
the Clifford formalism a particular Barut--Zanghi (BZ) solution, which 
refers (at the classical limit) to an ``internal" helical motion with a 
{\em time--like} speed [and is 
here shown to originate from the superposition of positive and negative
frequency 
solutions of the Dirac equation].\\
\h Then, we show how to construct solutions of the Dirac equation describing
helical motions with {\em light--like} speed, which meet very well the standard
interpretation of the velocity operator in the Dirac equation theory (and agree
with the solution proposed by Hestenes, on the basis ---however---
of ad-hoc assumptions that are unnecessary in the present approach).\\
\h The above results appear to support the conjecture that the Zitterbewegung 
motion (a helical motion, at the classical limit) is responsible
for the electron spin.

\vfill
 
\newpage

{\bf 1.} {\em Introduction -- \ }  The mysterious Zitterbewegung motion,
associated since long with the electron structure, seems to be responsible
for the electron spin.  Indeed, Schroedinger$^{1}$ proposed the
electron spin to be a consequence of a local circulatory motion, constituting
the Zitterbewegung (zbw) and resulting from the interference between 
positive and negative energy
solutions of the Dirac equation. Such an issue turned out to be of renewed
interest, following recent work, {\em e.g.}, by Barut {\em et al.},$^{2-4}$
Hestenes,$^{5,6}$ and Pav\v{s}i\v{c} {\em et al.}$^{7}$.  \ Let us recall 
that the pair of conjugate variables $(x^{\mu},
p_{\mu})$ is not enough to characterize the {\em spinning} particle.  In fact, 
after introducing the additional classical spinor variables 
($z,i\bar{z}$) \ [where $z$ is a Dirac spinor; and \ ${\bar z} \equiv z^{\dar} 
\gabf^{0}$], \ Barut and Zanghi$^{2}$
associated the electron spin and  zbw ---at the classical
limit--- with a canonical system [a point $\cal Q$ moving along a
cylindrical helix] which after quantization describes the Dirac 
electron.$^{4}$. \ In ref.$^{7}$, Pav\v{s}i\v{c} {\em et al.} presented a
thoughtful study of the above results in terms of the Clifford algebra
formalism; however, they$^{7}$ left many questions still open, and aim of 
this note is addressing a few of them.\\
\h First of all, let us recall that Hestenes' analysis$^{5}$ was based on his
reformulation$^{5,6,8,9}$ of Dirac theory in terms of the so-called ``Clifford
space--time algebra (STA)" ${\R}_{1,3}$. \ For details about the Dirac--Hestenes
(DH) spinors and the Clifford bundle formalism (used also 
below), see {\em e.g.} refs.$^{10-14}$. \ In Hestenes' papers$^{5,6}$ on his 
``zbw interpretation
of quantum mechanics", an ad-hoc assumption appeared, when he identified the
electron velocity with the light--like vector \ $u = e_0 - e_2$, \ with \ 
$e_{\rm i} = \psi \ga_{\rm i} \psit$, \ [${\rm i}=0,2$], \ where $\psi$ is a
(plane--wave) DH spinor field$^{5,6,10-12}$ satisfying the Dirac--Hestenes
equation ({\em i.e.}, the equation representing the ordinary Dirac equation
in the Clifford formalism). Then, he represented the electron internal 
structure by a light--like helical {\em motion} 
of a sub--microscopic ``constituent"
$\cal Q$, such that (for a suitable choice of the helix parameters) the helix 
diameter equals the electron Compton wavelength
and the angular momentum of the zbw yields the correct electron spin. \ At last,
he directly associated the complex phase factor of the electron wave--function
with the zbw motion.\\
\h We are going to show, among other things, that Hestenes' ad--hoc assumption,
namely that $u = e_0 - e_2$, is not necessary. \ More in general, below we
shall show:\\
(i) \ {\em how} to construct in the center-of-mass (CM) frame a particular 
solutions of the DH equation which correspond
(at the classical limit) to a
helical motion with time--like velocity. \ [It will result to be 
superpositions$^{(\# 1)}$ of positive 
\footnotetext{$^{(\# 1)}$ Not everybody may like the appearance of such a 
superposition. \ It should be clearly noticed, however, that in a generic frame
our 
``helical" wave-functions have to be expressed as superpositions of positive and
negative frequency solutions (of the Dirac equation) only when we want to
regard themselves  ---as done in this paper---  as solutions of the {\em Dirac}
equation (17). On the contrary, when we regard those ``helical" wave-functions
as solutions of our new, non-linear, Dirac--like equation (6'), no 
superposition of that kind is needed: cf. ref.$^{7}$; this can be considered 
a further point in favour of eq.(6').}
and negative energy solutions of the DH equation; that is to say (as thoroughly 
explained in refs.$^{15}$, only on the basis of relativistic classical 
physics), 
superpositions of particle and antiparticle solutions of the Dirac equation.
\ This suggests, as already pointed out in ref.$^{7}$, the BZ theory to be 
indeed equivalent in the CM frame to Dirac's theory;\\
(ii) \ that there exist also solutions of the DH equation that correspond 
at the classical limit to a helical
path with the light speed $c$, in which case the velocity operator$^{16}$ (as 
expected in ref.$^{5}$) {\em can} actually be identified with $u = e_0 - e_2$.\\

\vspace*{1.0 cm}

{\bf 2.} {\em Spin and electron structure -- \ }  Let us start by recalling 
that ---as shown by us in ref.$^{7}$--- the
dynamical behaviour of a spinning point-like particle, that follows a 
world--line \ $\sig = \sig (\tau)$, \ must be individuated ---besides by
the canonical variables $(x^{\mu}, p_{\mu})$--- also by the 
{\em Frenet tetrad}$^{\: 7,17}$

\

\hfill{$e_{\mu} \: = \: R \ga_{\mu} {\til{R}} \: = \: \Lambda_{\mu}^{\nu} 
\ga_{\nu} \: ; \;\;\;\;\;\; \Lambda_{\mu}^{\nu} \in L_{+}^{\upa}$ \hfill} (1)

\

where $e_{0}$ is parallel to the particle velocity $v$ (even more, $e_0 = v$ 
when we can use as parameter $\tau$ the particle proper--time). \ In eq.(1), 
the 
tilde denotes the reversion operation in the STA; namely: \ 
$\widetilde{AB} = \til {B} \til {A}$, \ and \ $\til {A} = A$ \ if $A$ is a
scalar or a vector, while \ $\til {F}=-F$ \ when $F$ is a 2-vector. \ Quantity
 \ $R = R(\tau)$ is a ``Lorentz rotation"$^{\, 18}$ \ [more precisely, \ 
$R \: \in \: 
{\rm Spin}_{+}(1,3) \simeq {\rm SL}(2,\C)$, \ and a Lorentz transform of 
quantity $a$ is given by $a' = R a \til{R}$]. \ Moreover \ $R \til{R} \: = \: 
\til{R} R \: = \: 1 \:$. \ The Clifford STA fundamental
unit--vectors $\ga_{\mu}$, incidentally, should not be confused with the Dirac 
{\em matrices} $\gabf_{\mu}$. \
Let us also recall that, while the orthonormal vectors $\ga_{\mu} \equiv 
{\pa / {\pa x^{\mu}}}$ constitute a {\em global} tetrad in Minkowski 
space-time 
(associated with a given inertial observer), on the contrary the Frenet
tetrad $e_{\mu}$ is defined only along $\sig$, in such a way that $e_0$
is tangent to $\sig$. \ At last, it is: \ $\ga^{\mu} = 
\eta^{\mu \nu} \ga_{\nu}$, \ and \ $\gamma_5 \equiv \ga_0 \ga_1 \ga_2
\ga_3$ \ is the volume element of the STA.\\
\h If \ $\Psi_{\rm D} \in \C^{4}$ \ is an ordinary Dirac spinor, in the STA it
will be represented by

\
                                                                               
\hfill{$\Psi_{\rm D} \longra \Psi \: = \: \psi \veps \: \in \: 
{\R}_{1,3} \; ,$ 
\hfill} (2)

\

where \ $\psi \in {\R}^{+}_{1,3}$ \ is called a Dirac--Hestenes spinor$^{8,14}$
and $\veps$ is an appropriate primitive idempotent of ${\R}_{1,3}$. \ It is 
noticeable that the spinor field $\psi$ carries {\em all} the essential 
information
contained in $\Psi$ (and $\Psi_{\rm D}$, and ---when it is nonsingular,
$\psi \psit \neq 0$--- it admits$^{8,9}$ a remarkable canonical decomposition 
in terms of a Lorentz rotation $R$, a duality transformation$^{19}$ \ 
$\erm^{{\bt \gamma_{5}} /2}$, \ and a dilation $\sqrt{\rho}$:

\

\hfill{$\psi \; = \; \r^{1 \over 2} \erm^{{\bt \gamma_{5}} /2} R \; .$ 
\hfill} (3)

\

In eq.(3), the normalization factor $\rho$ belongs to $\R^{+}$; \ quantity
$\bt$ is the Takabayasi angle;$^{20}$ \ and \ $\erm^{{\bt \gamma_{5}}} = 
+1$ \ for the electron (and $-1$ for the positron). \ \ Then, the Frenet
tetrad can also be written:

\

{\hfill{$\rho e_{\mu} \; = \; \psi \ga_{\mu} \psit \; .$
\hfill} (4)

\

\h Now, let us take as the lagrangian for a classical spinning particle, 
interacting with the electromagnetic potential $A$ (a 1--vector) the
expression$^{7}$

\

\hfill{$\Lc \; = \; \lan \psit \dopsi \ga_1 \ga_2 \: + \: p(\dox - 
\psi \ga_0 \psit) \: + \: eA \psi \ga_0 \psit \ran_0 \; ,$
\hfill} (5)

\

which is the translation$^{7}$ of the BZ lagrangian$^{2}$ into the Clifford 
bundle formalism [cf. also ref.$^{21}$].  \ In eq.(5), 
\ $\lan \;\;\; \ran_0$ means ``the scalar part" 
of the Clifford product; the dot represents the derivation with respect to 
the invariant time--parameter $\tau$; and $p$ can be regarded as a 
Lagrange multiplier. Let us note, moreover, that the BZ theory is also a
hamiltonian system, as shown in refs.$^{22,23}$ by means of Clifford
algebras. \ Then,
the Euler--Lagrange equations yield a system of three independent equations:

\

\hfill{$\dopsi \ga_1 \ga_2 + \pi \psi \ga_0 \; = \; 0 $
\hfill} (6)

\hfill{$\dox \; = \; \psi \ga_0 \psit $          
\hfill} (7)

\hfill{$\dopi \; = \; e F \cdot \dox \; , $
\hfill} (8)                                                

\
 
where $\pi \equiv p - eA$ \ is the kinetic momentum; \ $F \equiv \pa \wedge A$ 
\ is the electromagnetic field (a bivector, in Hestenes' language); \  
$\dpar \: = \: \ga^{\mu} \dpar_{\mu}$ is the Dirac operator; \ and symbols
$\cdot$ and $\wedge$ denote the internal and external product, respectively, 
in the STA. \ The system (6)--(8) is just that one appeared in ref.$^{2}$,
but written$^{7}$ in terms of the STA language.\\ 
\h Let us pass to the free case, \ $A = 0$, \ for which one gets
 
\

\hfill{$\dopsi \ga_1 \ga_2 + p \psi \ga_0 \; = \; 0 $
\hfill} (9)

\hfill{$\dox \; = \; \psi \ga_0 \psit $          
\hfill} (10)

\hfill{$\dot p \; = \; 0 \; .$
\hfill} (11)                                                

\

If we choose the {$\ga_{\mu}$} frame in such a way that

\

\hfill{$p \; = \; m \, \ga_0$
\hfill} (12)

\

is a constant vector in the $\ga_0$ direction, with \ $p^2 = m^2$, \ then 
for the system (9)--(11) we find the {\em solution}: 

\

\hfill{$\psi (\tau) \; = \; \cos (m\tau) \, \psi (0) \: + \: \sin (m\tau) \,
\ga_0 \psi (0) \, \ga_0 \ga_1 \ga_2 \; ,$
\hfill} (13)

\

where $\psi (0)$ is a constant spinor, which translates into the Clifford
language the solution$^{2}$ found by BZ for their analogous system of
equations. \ In the case of solution (13), it holds:

\

\hfill{$v(\tau) \; \equiv \: {\dot x}(\tau) \; = \;  \dis{pH \over m^2} \:
+ \: [v(0) - \dis{pH \over m^2}] \, \cos (2m\tau) \: + \: \dis{{\dot v}(0) \over
2m} \, \sin (2m\tau) \; ,$
\hfill} (14)

\

which clearly shows the presence of an internal helical motion ({\em i.e.}, 
at the classical level, of the zbw phenomenon).\\ 
\h In eq.(14) we have \ $H = v \cdot p = \:$constant. \  If the constant is 
chosen to be $m$:            

\

\hfill{$H \; = \; p \cdot v \; = \; m \; ,$
\hfill} (15)

\

and ---more important--- if now \ $\psi (x)$ \  is a DH spinor {\em field}
such that its {\em restriction} to the world--line $\sig$ yields \ $\psi
(\tau)$, \ namely \ $\psi_{\mid \sig}(x) = \psi (\tau)$, \ {\em then}
eq.(13) writes

\

\hfill{$\psi (x) \; = \; \cos (p \cdot x) \: \psi(0) \: + \:  \sin
(p \cdot x) \: \ga_0 \, \psi(0) \, \ga_0 \ga_1 \ga_2 \; ,$
\hfill} (13')

\

which now is a quantum wave-function, solution, as we are going to see, of
the Dirac equation! \ In fact, it is:

\

\hfill{$\dot{\psi} \; \equiv \; {\dis{{{\drm \psi} \over {\drm \tau}}}} \; = \; 
v^{\mu} \pa_{\mu} \psi \; = \; (v \cdot \pa) \, \psi \; ,$
\hfill} (16)

\

and, for any eigen-spinors $\psi$ of \ $\hat{p} \psi \equiv \pa \psi \, \ga_1
\ga_2 = p \psi$, \ one gets$^{7}$ by using the equalities 
 \ $(v \cdot \pa) \psi \ga_1 \ga_2 \, = \, (v \cdot \hat{p}) \psi \, = \,
(v \cdot p) \psi \, = \, m \psi$ \  
that eq.(9) transforms into the 
ordinary {\em Dirac equation} in its Dirac--Hestenes form:$^{5-8}$

\

\hfill{$\pa \psi \ga_1 \ga_2 \: + \: m \psi \ga_0 \; = \; 0 \; .$
\hfill} (17)  

\

\h Notice once more that, while eq.(9) refers to $\psi = \psi(\tau)$, on the
contrary the quantum equation (17) refers to the spinor field 
\ $\psi = \psi(x)$ \ [such that \ 
$\psi_{\mid \sig}(x) = \psi (\tau)$]. \ Then, eq.(13') is an actual solution
of eq.(17); while eq.(13) ---if you want--- can be said to be a solution of
eq.(17) {\em when} this equation is {\em restricted} along the
stream--line $\sig$ ({\em i.e.}, the world-line of the ``sub-microscopic"
object $\cal Q$). \ When moving from the classical to the quantum 
interpretation, bearing in mind the Feynman paths formalism,
one has to pass from considering a single helical path to
consider a {\em congruence} of helical paths.$^{\# 2}$\\
\footnotetext{$^{\# 2}$ Such a congruence of helical paths can be 
regarded as constituting a quantum ``fluid"; in such a fluid, however, we 
would have
a flux of energy--momentum and angular momentum also along the normal to the
velocity stream--lines. Therefore, this fluid would be a Weyssenhoff 
fluid,$^{24}$ rather than a ``Dirac fluid".$^{23}$}
                                 
\h Let us go back, for a moment, to the system (6)--(9). We may notice that 
---using eq.(16)--- the [total derivative] equation (6) can be rewritten$^{7}$ 
in the [partial derivative] noticeable form:

\

\hfill{$(\psi \ga_0 \psit) \cdot \pa \psi \ga_1 \ga_2 + \pi \psi \ga_0 \; 
= \; 0 \; ,$
\hfill} (6')

\
 
which is a {\em non-linear}, Dirac--like equation.\\ 

\vspace*{1.0 cm}

{\bf 3.} {\em Time--like helical motions -- \ }  Let us prove, 
now, that
solution $\psi (x)$ of eq.(17), which reduces to the $\psi(\tau)$ given by
eq.(13) on the world--line $\sig$ of the point-like constituent $\cal Q$, 
is indeed
a superposition$^{(\# 1)}$ of positive and negative energy states, {\em i.e.}, 
of particle and antiparticle states (for a purely kinematical reinterpretation 
of the negative energy states in terms of antiparticles, without any recourse 
to a ``Dirac sea", cf. refs.$^{15}$). \ Indeed, $\psi(\tau)$ can be
written in the CM frame as:

\

\hfill{$\psi(\tau) \; = \; {1 \over 2} [\psi(0) + \ga_0 \psi(0) \ga_0] \:
\exp (\ga_1 \ga_2 m \tau) \: + \: {1 \over 2} [\psi(0) - \ga_0 \psi(0) \ga_0] \:
\exp (- \ga_1 \ga_2 m \tau) \; .$
\hfill} (18)                      

\

But quantities

\

\hfill{${1 \over 2} [\psi \pm \ga_0 \psi \ga_0] \; \equiv \; \La_{\pm} (\psi)$
\hfill} (19)

\

are nothing but the positive and negative energy 
{\em projection operators}$^{\, 14}$ \ $\La_{+}, \: \La_{-} \in {\rm End}
(\R_{1,3})$, \ respectively; so that eq.(18) can read

\

\hfill{$\psi(\tau) \; = \; \psi_{+}(0) \: \exp (\ga_1 \ga_2 m \tau) \: + \: 
\psi_{-}(0) \: \exp (- \ga_1 \ga_2 m \tau) $
\hfill} (20)                      

\

where \ $\psi_{\pm}(0) \equiv \La_{\pm}[\psi(0)]$, \ which proves our claim. \
It follows that any $\psi(x)$, such that \ $\psi_{\mid \sig}(x) = \psi (\tau)$,
 \ is a solution of the DH equation (17) in the CM frame.\\
\h It is  clear that one can now construct ``helical" solutions of the DH
equation with time--like velocity, that manifest a zbw phenomenon. \ Quantity
$\tau$, let us repeat, is here an invariant time--parameter. \ 
Now, to construct a time--like solution, let us take as a concrete example:

\

\hfill{$\psi(0) \equiv \sqrt{\r_{+}} \, + \, \sqrt{\r_{-}} \, \ga_1
\ga_0 \; ; \;\;\;\;\;\; \sqrt{\r_{\pm}} \, \equiv \, 
\sqrt{\r_{\pm}}(0) \; .$
\hfill} (21)  

\

Since

\

\hfill{$\psi(0) \: \psit(0) \; = \; \r_{+} - \r_{-}$
\hfill} (22)

\

we can put, for simplicity, \ $\r_{+} - \r_{-} \, = \, 1$. \ In this case, \
$\psi_{+}(0) = \sqrt{\r_{+}}$ \ and \ $\psi_{-}(0) = 
\sqrt{\r_{-}} \, \ga_1 \ga_0$. \ \ Then, by using eq.(14), from eq.(21)
it follows that

\

\hfill{$v(0) \; = \; (\r_{+} + \r_{-}) \ga_0 \, + \, 2 \, \sqrt{\r_{+} \r_{-}}
\: \ga_1 \: ; \;\;\;\; \dot{v}(0) \; = \; 4 m \, \sqrt{\r_{+} \r_{-}} \: \ga_2$
\hfill} (23)

\hfill{$H \; = \; m \, (\r_{+} + \r_{-}) \; ,$
\hfill} (24)

\

and we end up with:

\

\hfill{$v(\tau) \; = \; (\r_{+} + \r_{-}) \ga_0 + 2 \, \sqrt{\r_{+} \r_{-}} \:
[\ga_1 \, \cos (2m\tau) + \ga_2 \, \sin (2m\tau)] \; ,$
\hfill} (25)

\

for which it is \ $v^2(\tau) = 1$ \ (so that we got a special 
{\em time--like} case).  \ For
the spin bivector \ $S = {\hbar \over 2} \psi \ga_2 \ga_1 \psit$ \ and the
spin vector \ $s = {\hbar \over 2} \psi \ga_3 \psit$, \ we have in this case  

\

\hfill{$S \; = \; {\dis{{1 \over 2}}} (\r_{+} + \r_{-}) \ga_2 \ga_1 \: + \:
\sqrt{\r_{+} \r_{-}} \: [\ga_0 \ga_1 \sin (2m\tau) - \ga_0 \ga_2 \cos
(2m\tau)]$ 
\hfill} (26)

\hfill{$s \; = \; {\dis{{1 \over 2}}} \, \ga_3 \; ; \;\;\;\;\;\;\;\;\;\;
[\hbar = 1] \; .$
\hfill} (27)

\

The velocity \ $v(\tau) \equiv {\drm x}/{\drm \tau}$, \ with $x(\tau) = 
x^{\mu}(\tau) \ga_{\mu}$, \ is easily integrated, from eq.(25), to give: 

\

\hfill{$x(\tau) \; = \; (\r_{+} + \r_{-}) \tau \ga_0 \: + \: 
{\dis{{\sqrt{\r_{+} \r_{-}}\over m}}} \: [\ga_1 \sin (2m\tau) - \ga_2 \cos
(2m\tau)] \: + \: x_0 \; ,$
\hfill} (28)

\

which is the parametric equation of a {\rm helix}, whose diameter is \ $D =
2m \sqrt{\r_{+} \r_{-}}$. \ Equation (24) suggests to introduce a  
renormalized mass \ $M \equiv m (\r_{+} + \r_{-})$. \ If one assumed the
maximum diameter of that helix to be the electron Compton wave--length,
one would get for the new mass $M$ the upper limit \ $M = m \sqrt{2}$.\\
\h It is worth observing that, from eq.(28), for \ $L \equiv x \wedge p$ \
we have:

\

\hfill{$L \; = \; \sqrt{\r_{+} \r_{-}} \: [\ga_1 \ga_0 \sin (2m\tau) \, -
\, \ga_2 \ga_0 \cos (2m\tau)] \; ,$
\hfill} (29)

\

where we neglected the constant contribution $m x_0 \wedge \ga_0$. \ Notice
that \ ${\dot L} \neq 0$, \ so that $L$ alone is not conserved; \ however,
in view of eq.(26), we obtain that the {\em total} angular momentum $J$ 
{\em is} conserved:

\

\hfill{$J \; \equiv \; L + S \; = \; {\dis{1 \over 2}} \, (\r_{+} + \r_{-})
\, \ga_2 \ga_1 \; ; \;\;\;\;\;\;\; {\dot J} \; = \; 0 \; ;$
\hfill} (30)

\

which implies a {\em nutation} of the spin plane. \ \ 
Under the above assumption (that the maximum helix diameter be the electron 
Compton wave--length), one would get \ 
 $\mid J \mid \, \equiv \, [J \cdot 
{\til J}]^{1/2} \, \leq \, {\sqrt{2} \over 2}$.\\

\vspace*{1.0 cm}

{\bf 4.} {\em Light--like helical 
motions -- \ }   Finally, let us show how one can obtain solutions of the
BZ theory with speed of the helical motion equal to the light speed $c$. \
To this aim, it is enough to choose \ $\r_{+} = \r_{-} = 1/2$, \ so that \
$\r_{+} - \r_{-} = 0$. \ [In this case, \ $\psi (0) = (1 + \ga_1 \ga_0)/
\sqrt{2}$ \ is a singular spinor, actually a Majorana spinor since the
charge conjugation operator C is such$^{14}$ that \ ${\rm C} \, \psi =
\psi \ga_1 \ga_0 \:$].\\
\h In fact, from eq.(14) one then gets

\

\hfill{$v(0) \; = \; \ga_0 \, + \, \ga_1 \: ; \;\;\;\;\;\;\; \dot{v}(0) \; = \;
2 m \ga_1 $
\hfill} (31)

\hfill{$H \; = \; m \; ,$
\hfill} (32)

\

which yield for the velocity $v$:

\

\hfill{$v(\tau) \; = \; \ga_0 \: + \: \ga_1
\, \cos (2m\tau) \: + \: \ga_2 \, \sin (2m\tau) \; ,$
\hfill} (33)

\
 
in which it is now \ $v^2(\tau) = 0$ \ ({\em light--like} case). \ One can see
that, after a convenient rotation, it is possible to write $v$ as
 \ $v = e_0 - e_2$, \  like in Hestenes' papers.$^{5,6}$ \ \ Moreover, we now
have:
 
\

\hfill{$S \; = \; {\dis{{1 \over 2}}} \ga_2 \ga_1 \: + \:
{\dis{{1 \over 2}}} \, [\ga_0 \ga_1 \sin (2m\tau) - \ga_0 \ga_2 \cos
(2m\tau)]$ 
\hfill} (34)

\hfill{$s \; = \; {\dis{{1 \over 2}}} \, \ga_3 \; ; \;\;\;\;\;\;\;\;\;\;
[\hbar = 1] \; .$
\hfill} (35)

\

Integration of eq.(33), now, yields:

\

\hfill{$x(\tau) \; = \; \tau \ga_0 \: + \: {\dis{{1 \over 2m}}} \: 
[\ga_1 \sin (2m\tau) - \ga_2 \cos (2m\tau)] \: + \: x_0 \; ,$
\hfill} (36)

\

and one can verify that in this case the helix diameter is actually the 
Compton wave--length of the electron! \ For  $L \equiv x \wedge p$ \
we obtain again

\

\hfill{$L \; = \; {\dis{1 \over 2}} \, [\ga_1 \ga_0 \sin (2m\tau) \, -
\, \ga_2 \ga_0 \cos (2m\tau)] \; ; \;\;\;\;\;\; {\dot L} \neq 0 \; ,$
\hfill} (37)

\

whilst the conserved quantity is

\

\hfill{$J \; \equiv \; L + S \; = \; {\dis{1 \over 2}} 
\, \ga_2 \ga_1 \; ; \;\;\;\;\;\;\; {\dot J} \; = \; 0 \; ;$
\hfill} (38)

\

which again implies a nutation of the spin plane.\\
\h In conclusion, we showed how to construct in the CM frame 
solutions of the Dirac equation associated (at the classical limit) to a helical
motion with the light speed $c$. \ And, in particular,
we got results quite similar to Hestenes' without his ad--hoc 
assumptions$^{5}$ / [by making recourse, however, to a superposition$^{\# 1}$
of particle and antiparticle$^{15}$ solutions of the DH equation. \ Indeed, 
the part of the velocity $v(\tau)$
responsible for the zbw, namely \ $\sqrt{\r_{+} \r_{-}} \,
[\ga_1  \cos (2m\tau) + \ga_2  \sin (2m\tau)]$, \ is given by an interference 
term \ $\psi_{+} \ga_0 \psit_{-} + \psi_{-} \ga_0 \psit_{+}$, \ in which 
quantities \ $\psi_{+} = \sqrt{\r_{+}} \, \exp (\ga_1 \ga_2 m \tau)$ \ and
 \ $\psi_{-} = \sqrt{\r_{-}} \, \ga_1 \ga_0 \, \exp ( - \ga_1 \ga_2 m \tau)$
 \ are the positive--energy (particle) and negative--energy 
(antiparticle$^{15}$) states].\\

\vspace*{1.5 cm}

{\bf 5.} {\em  Acknowledgements -- \ }  The authors are grateful for 
discussions to A. Bugini, R. Garattini, L. Lo Monaco, G.D. Maccarrone,
J.E. Maiorino, M. Pav\v{s}i\v{c}, R. Pucci,
F. Raciti,  
M. Sambataro, Q.A.G. de Souza, and particularly to E.C. de Oliveira, M. Pauri 
and S. Sambataro.

\end{document}